\documentclass[fleqn,usenatbib]{mnras}

\usepackage{newtxtext}

\usepackage[T1]{fontenc}

\DeclareRobustCommand{\VAN}[3]{#2}
\let\VANthebibliography\thebibliography
\def\thebibliography{\DeclareRobustCommand{\VAN}[3]{##3}\VANthebibliography}

\usepackage{graphicx}	
\usepackage{amsmath}	
\usepackage{amssymb}	

\newcommand{\krome}{\textsc{krome}}

\newcommand{\msun}{{\rm M_\odot}}

\graphicspath{{Pictures/}}

\title[Chemical post-processing]{Chemical post-processing of magneto-hydrodynamical simulations of star-forming regions: robustness and pitfalls}

\author[S. Ferrada-Chamorro et al.]{
Sim\'on Ferrada-Chamorro,$^{1}$\thanks{Now at IPAG, Université Grenoble Alpes, 38000 Grenoble, France; Email:  \href{mailto:simon.ferrada@univ-grenoble-alpes.fr}{simon.ferrada@univ-grenoble-alpes.fr}}
Alessandro Lupi,$^{2}$
and Stefano Bovino$^{1}$
\\
$^{1}$Departamento de Astronom\'ia, Universidad de Concepci\'on, Barrio Universitario, Concepci\'on, Chile\\
$^{2}$Dipartimento di Fisica ``G. Occhialini'', Universit\`a degli Studi di Milano-Bicocca, Piazza della Scienza 3, I-20126 Milano, Italy\\
}

\date{Accepted XXX. Received YYY; in original form ZZZ}

\pubyear{2020}

\begin{document}
\label{firstpage}
\pagerange{\pageref{firstpage}--\pageref{lastpage}}
\maketitle
\begin{abstract}
A common approach to model complex chemistry in numerical simulations is via post-processing of existing magneto-hydrodynamic simulations, relying on computing the evolution of chemistry over the dynamic history of a subset of particles from within the raw simulation. Here, we validate such a technique, assessing its ability to recover the abundances of chemical species, using the chemistry package \textsc{krome}. We also assess, for the first time, the importance of the main free input parameters, by means of a direct comparison with a self-consistent state-of-the-art simulation in which chemistry was directly coupled to hydrodynamics. We have found that the post-processing is highly reliable, with an accuracy at the percent level, even when the most relaxed input parameters are employed. In particular, our results show that the number of particles used does not affect significantly the average properties, although it suppresses the appearance of possibly important spatial features. On the other hand, the choice of the integration time-step plays a crucial role. Longer integration time-steps can produce large errors, as the post-processing solution will be forced towards chemical equilibrium, a condition that does not always necessarily apply. When the interpolation-based reconstruction of chemical properties is performed, the errors further increase up to a factor of $\sim2$. Concluding, our results suggest that this technique is extremely useful when exploring the relative quantitative effect of different chemical parameters and/or networks, without the need of re-running simulations multiple times, but some care should be taken in the choice of particles sub-sample and integration time-step.
\end{abstract}




\section{Introduction}
Since the detection of the first molecules in the interstellar medium (ISM; \citealt{swings1937,douglas1941}), there has been an increasing interest in the study of chemistry in astronomical contexts, up to the point that, nowadays, the study of both chemical structure and evolution of any given system under scrutiny is widely accepted as crucial to completely understand it. Unfortunately, the inclusion of self-consistent chemical evolution in theoretical models in the aim at further constraining the processes happening in astronomical environments is very complicated. Chemical reactions work on very short time-scales compared to other dynamical processes, and, on top of this, they strongly depend on the local properties (density and temperature) of the medium and the impinging radiation flux. 

From a numerical perspective, an accurate treatment of any system evolution as a whole would require on-the-fly (OTF) non-equilibrium chemistry calculations in hydrodynamic simulations, especially when the chemical state of the gas affects the thermodynamics of the system, for instance on galactic scales \citep[see, e.g.][]{2014MNRAS.440.3349R,2016A&A...590A..15B}. While most of the galaxy-scale studies to date are still based on the post-processing of the simulations, either via pre-computed \textsc{cloudy} \citep{ferland1998} tables \citep{pallottini2017b,olsen2017,pallottini2019,katz2019,arata2020}, or determined from chemical networks under the assumption of photo-ionisation equilibrium \citep{keating2020}, the last decade has seen an increasing effort by several groups to properly include non-equilibrium chemistry in simulations, from simple models tracing primordial species \citep{gnedin2009,christensen2012,tomassetti2015,katz2017,pallottini2017b,lupi2018,nickerson2018} to more accurate ones that also include metal elements \citep{Glover2010,Hu2016, richings2016,capelo2018,lupi2020a,lupi2020b}.  

Conversely to galactic scales, the chemistry complexity exponentially grows when simulating the small scales typical of pre-stellar cores and proto-planetary disks. During the initial phases of the star formation process, pre-stellar cores are characterised by extremely high densities ($n_{\rm H_2} \gtrsim 10^4$ cm$^{-3}$) and low temperatures ($T<20$ K), and exhibit a well-defined chemical structure determined by characteristic chains of chemical reactions \citep{caselli2012,2021PhR...893....1O}. These conditions are ideal for the adsorption of heavy species on the surface of dust grains, a process known as freeze-out \citep{bacmann2002,bacmann2003,2006A&A...455..577T}, and thus a plethora of gas-grain interactions and reactions on the surfaces of such grains starts to take place, further increasing the chemical complexity of these objects \citep[see, e.g.,][]{2018IAUS..332....3V}. As a consequence, chemical networks exponentially grow, making on-the-fly calculations in already expensive 3D magneto-hydrodynamic (MHD) simulations computationally unfeasible. Relevant examples of still unresolved features due to the lack of robust and consistent chemical modelling are, for instance, the high-abundances of NH$_3$ observed towards the central regions of pre-stellar cores \citep{2004A&A...416..191T,crapsi2007,sipila2019}, or the convoluted processes governing the formation and evolution of interstellar complex organic molecules (iCOMs) \citep{jimenez-serra2016,punanova2018}.

The need to include gas-grain interaction and surface chemistry in the study of star-forming regions becomes then fundamental for the complete understanding of particular features and chemical complexities observed towards these objects. The common approach, in these cases, is to build low-dimensionality models, which are qualitatively useful to explore the different formation paths of a given iCOM or to perform sensitivity studies, by mapping a large parameter space. To date, there are no works where the formation of iCOMs is consistently followed in three-dimensional simulations, and the available studies rely instead upon the post-processing of the simulation outputs with a non-equilibrium chemistry solver \citep[see, e.g.][]{ruaud2018,coutens2020}. This approach is generally based on the selection of a sub-sample of resolution elements (particles) from standard pure MHD simulations, i.e. performed without OTF chemistry, and on the subsequent integration of the chemical equations on the already pre-determined dynamical history of said particles. On the other hand, \citet{priestley2018,priestley2019} suggested an even more convenient post-processing approach based on the parametrisation of the density structure of a collapsing pre-stellar core as a function of both time and radius, facilitating the inclusion of more complex dynamical processes, such as ambipolar diffusion, which become more important during the process of star formation. However, forcing the electron density to change as the solely function of dynamics without including any chemistry, as done by \citet{priestley2019}, can introduce large errors in the subsequent post-processing analysis.

Another important limitation is that the aforementioned techniques are usually applied to isothermal simulations. When the thermodynamics of the gas is taken into account, indeed, a minimal network of chemical reactions should always be included, to guarantee the correct thermal behaviour, which in turns depends on key species abundances (coolants), as for instance C$^+$ and CO. In these cases, it is unavoidable to couple OTF chemistry with dynamics. As an example, a minimal network able to follow CO formation/destruction and to give the right thermal behaviour in molecular cloud simulations, already requires $\sim$40 species and hundreds of reactions \citep[see e.g.][]{2017MNRAS.466.1259G}. Once the thermal behaviour is correctly assessed, post-processing can be still pursued to include more complex processes, like the formation of specific iCOMs, or isotopologues and isomers of specific tracers. 

Overall, post-processing techniques have the ability to explore the impact of different sources of chemical degeneracy on the evolution of specific tracers, without incurring in computationally intensive problems, but, despite the different results obtained so far, a proper assessment of their reliability and limitations, performed against self-consistent MHD simulations with OTF chemistry, is still missing.  With this in mind, we assess here the reliability of post-processing chemistry in 3D MHD simulations in the context of star-forming collapsing cloud fragments, by directly comparing it to the results obtained by MHD simulations including coupled OTF non-equilibrium chemistry calculations. Thanks to the latter, we have the unique opportunity to evaluate the uncertainties associated with the post-processing. For the first time, we can constrain parameters like the number of resolution elements in the sub-sample, numerical approximations in the interpolation schemes, and the time-discretisation for the dynamical properties update.

The paper is organised as follows. In Section~\ref{sec:methodology}, we briefly summarise the setup used and describe how chemistry calculations are performed in the post-processing of the simulation. The discussion of the validity and limitations of the post-processing approach relative to OTF chemistry calculations is presented in Sections~\ref{sec:validation} and~\ref{sec:comparison}, respectively. Finally, in Section ~\ref{sec:conclusions} we draw our conclusions.

\section{Methodology}
\label{sec:methodology}
Here, we describe the setup of the simulation considered in this work, providing a summary of the initial conditions, and the details of the chemical network employed, and then introduce the post-processing method we apply.

\subsection{The reference MHD simulation}
As a reference case for this work, we consider the slow-collapse `M1' core from the simulation suite presented in \citet{Bovino2019}, a 3D MHD simulation of the isothermal collapse of a high-mass pre-stellar core. The assumption of an isothermal gas helps us further facilitate the treatment of chemistry in post-processing, not requiring any `feedback' of the temperature resulting from the chemical calculations on to the gas dynamics. The simulation was performed with the MHD code \textsc{gizmo} \citep{hopkins2015,Hopkins2016}, using its meshless finite-mass method and the standard cubic-spline kernel with an effective number of neighbours of 32.

The initial conditions for the simulation consist of a $M_{\rm BE}=20\rm\, M_{\odot}$ core following a Bonnor-Ebert density profile \citep{Ebert1955,Bonnor1956} with core radius $R_{\rm BE} = 0.17$ pc and an homogeneous temperature $T = 15$ K. The average number density of the core is $\langle n \rangle \equiv (3M_{\rm BE})/(4\upi R_{\rm BE}^3) = 2.21\times 10^{4}$ cm$^{-3}$, corresponding to an average free-fall time of $t_{\rm ff} = 260$ kyr. The number of particles (from both the core and the background gas) was set to $6\times 10^5$, corresponding to a mass resolution of $3.33\times10^{-5}\,\msun$.

Alongside self-gravity, our simulation also includes ideal MHD, with the magnetic field aligned along the $z$-direction, and scaling with the perpendicular distance $R_\perp$ (in the $xy$ plane) as
\begin{equation}
    B_z(R_\perp)=B_0\left(1 + \frac{R_\perp}{R_{\rm BE}}\right)^{-\kappa},
\end{equation}
where B$_z$ is the component of the magnetic field parallel to the $z$-axis, $k=1.5$ an input scaling exponent, and $B_0$ the magnetic field strength at $R_\perp=0$. The latter is obtained by enforcing a user-defined mass-to-magnetic flux ratio $\mu\equiv M/\Phi = 2$, commonly used in simulations of pre-stellar cores \citep[see, e.g.][]{bastian2017,Kortgen2018,Goodson2016}. This choice yields an average magnetic field strength in our core of $\langle B \rangle = 46$ $\mu$G.

Finally, the core was assumed to be stirred by an initial turbulent velocity field following a Burgers-like power spectrum, with $E(k)\propto k^{-2}$ (see \citealt{bastian2017}). The turbulent field was normalised according to the assumed temperature $T$ and the desired Mach number $\mathcal{M} = 3$. The main parameters of the run are summarised on Table~\ref{tab:ICs1}.
\begin{table}
    \centering
    \caption{Fiducial values of important physical parameters used in the validation of the post-processing methodology.}
    \label{tab:ICs1}
    \begin{tabular}{ll}
    \hline Parameter & Value \\\hline
    \multicolumn{2}{c}{\textbf{Core properties}}\\
    Core mass $M_{\rm BE}$ & 20 M$_\odot$\\
    Core radius $r_{\rm BE}$ & 0.17 pc\\
    Average volume density $\langle n \rangle$ & 2.21$\times10^{4}$ cm$^{-3}$\\
    Mach number $\mathcal{M}$ & 3\\
    Virial parameter $\alpha_{\rm vir}$ & 4.32\\
    Average magnetic field $\langle B \rangle$ & 46 $\mu$G\\
    Free-fall time $t_{\rm ff}$ & 260 kyr\\
    \\
    \multicolumn{2}{c}{\textbf{Other parameters}}\\
    Temperature $T^{a}$ & 15 K \\
    CR ionisation rate per H$_2$ $\zeta_2$ & $2.5\times10^{-17}$ s$^{-1}$\\
    Average grain size $a$ & 0.035 $\mu$m \\
    Dust grain density $\rho_0$ & 3.0 g cm$^{-3}$ \\
    Dust-to-gas mass ratio $\mathcal{D}$ & $7.09\times10^{-3}$ \\
    Mean molecular weight $\mu$ & 2.4\\\hline
    \end{tabular}
    \\\flushleft $^a$ We assume $T_{\rm gas} = T_{\rm dust} = T$ throughout this work.
\end{table}

The chemical network employed in the simulation (and in our post-processing) is the same of \citet{Bovino2019,Bovino2020}. The network was created to explore the evolution of simple, yet fundamental C-N-O bearing tracers and deuterated species during the collapse of massive star-forming cores and clumps, following, for the first time, time-dependent freeze-out processes of key species in a 3D MHD setup. The species were initialised with the abundances reported in the second column of Table~\ref{tab:ICs2}, originally taken from \citet{sipila2015a}, but adapted to the molecular environment considered in our setup. In detail, species heavier than helium were no longer initialised in their atomic form, with the only exception being nitrogen, which was initialised half-atomic, half-molecular, due to the uncertainties about its main chemical state in dense environments. Our network contains mainly gas-phase chemistry, but also incorporates depletion and non-thermal desorption reactions for O, N, CO, and N$_2$. The adopted binding energies for these species are taken from \citet{Wakelam2017}, and are shown in the third column of Table~\ref{tab:ICs2}.
\begin{table}
    \centering
    \caption{Fiducial initial abundances $n$ of the species included in our network relative to the atomic hydrogen abundance $n_{\rm H}$. For the species that go through time-dependent depletion (see text), we also report the corresponding binding energy in the third column \citep[see][]{Wakelam2017}. Compared to the H$_2$ ortho-to-para ratio (OPR) usually assumed ($\leq0.1$), in this work we start from a more conservative value of 3.}
    \label{tab:ICs2}
    \begin{tabular}{lll}
    \hline Species & $n/n_{\rm{H}}$ & $E_{\rm B}$ [K] \\\hline
    oH$_2$ & $3.75\times10^{-01}$ & -- \\
    pH$_2$ & $1.25\times10^{-01}$ & -- \\
    HD & $1.50\times10^{-05}$ & -- \\
    He & $1.00\times10^{-01}$ & -- \\
    oH$_3^+$ & $1.80\times10^{-10}$ & -- \\
    pH$_3^+$ & $3.00\times10^{-09}$ & -- \\
    N & $1.05\times10^{-05}$ & 720 \\
    N$_2$ & $5.25\times10^{-06}$ & 1100 \\
    CO & $1.20\times10^{-04}$ & 1300 \\
    O & $1.36\times10^{-04}$ & 1600 \\
    GRAIN0 & $5.27\times10^{-11}$ & -- \\\hline
    \end{tabular}
\end{table}

\subsection{Post-processing hydrodynamic simulations with non-equilibrium chemistry}
\label{sec:postproc_method}
The computational cost of non-equilibrium chemistry in simulations poses significant limitations to its applicability, and evolving chemistry in post-processing can provide a viable cheaper alternative. Obviously, if all the particles in the simulation are used, we do not expect a large gain in terms of computational time, hence a more efficient solution is to only evolve a fraction of the gas resolution elements and then, if desired, interpolate the other elements according to the closest neighbour element in density (and/or any other relevant parameter) space.

\begin{figure}
    \centering
    \includegraphics[width=\columnwidth]{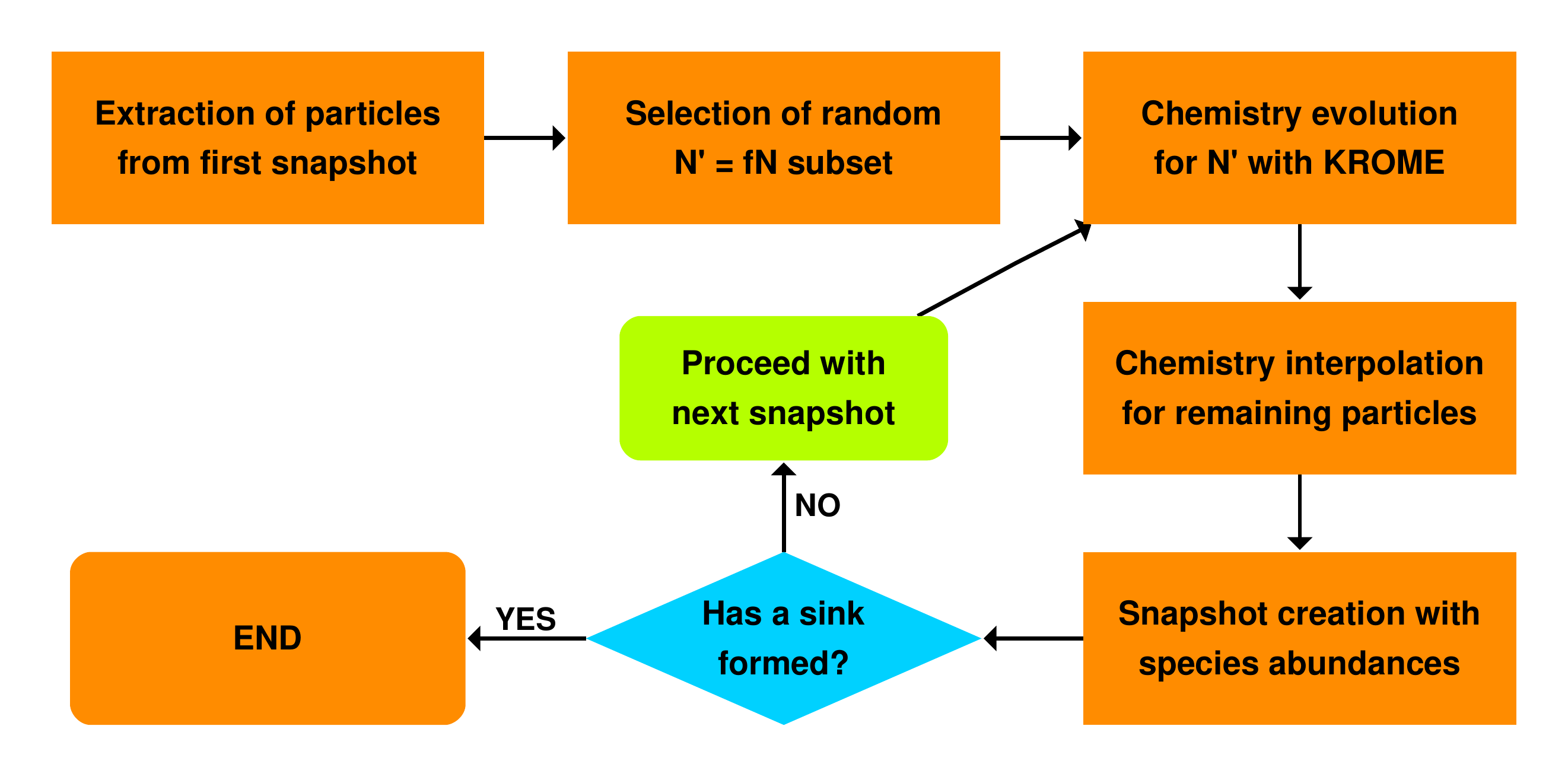}
    \caption{Schematic flowchart of the presented post-processing method.}
    \label{fig:flowchart}
\end{figure}

The post-processing method is based on the assumption that chemistry does not significantly affect the dynamical evolution of the system. While this is certainly true under the isothermal assumption made in this work, relaxing this condition would necessarily require on-the-fly chemistry with at least a minimal network to track the thermal evolution of the gas \citep[see, e.g.,][]{2017MNRAS.466.1259G}.  We thus start from a `standard' hydrodynamic simulation of our system, i.e. without chemistry, for which we collect the outputs in which the history of its dynamical properties is stored.
On these outputs, we then apply the non-equilibrium chemical evolution, following the procedure reported here, and also summarised in the scheme of Fig.~\ref{fig:flowchart}:
\begin{enumerate}
    \item we extract the density of every particle in the run from the simulation outputs, and build the corresponding density evolution history (by tracking the particle ID across multiple snapshots);
    \item we randomly select a subset $N'=f_{\rm part}N$ of the $N$ particles in the simulation initial conditions (i.e. $t=0$), with $f_{\rm part}\leq 1$ a user-defined parameter. In order to properly sample the entire dynamic range of the simulation, independent of the total number of particles used to map the pre-stellar core and the background, we also enforce a user-defined number fraction $\xi_{\rm core}$ to be extracted from the particles sampling the core and the remaining ones from the low-density background surrounding it;\footnote{In principle, $\xi_{\rm core}$ should be tuned to the actual number of particles employed to map the core and the background in the simulation, respectively. We notice that, neglecting $\xi_{\rm core}$, hence assuming an homogeneous sampling of the entire particle distribution, can result in the oversampling of the background, depending on the assumptions made in the simulations, as the minimum spatial resolution in the background region and the box size.}
    \item for the selected particles, we initialise the species abundances in the same way as the reference simulation with OTF chemistry, with the values reported in Table~\ref{tab:ICs2};
    \item we evolve the chemical abundances of the $N'$ subset with \krome, assuming a coarse integration time-step (i.e. the time at which the chemical abundances are actually stored) $\Delta t = m \Delta t_0$, with $\Delta t_0$ the time separation between the snapshots, and $m\geq 1$; over each integration step, the density of the particle is kept constant at the value obtained\footnote{We note that, although we assume a constant density at each step, different choices could be made, for example linearly interpolating between subsequent values, and the procedure would still be valid. However, such interpolation would simply represent an effective higher temporal resolution, and would require shorter $\Delta t$ for the chemistry solver, since each integration step necessarily assumes a constant gas density.} from the evolution history of every particle in the simulation at the beginning of each time-step;
    \item we estimate the species abundances for the remaining ($1-f_{\rm part}$) particles in the simulation from the closest post-processed particle in density-position space.
    In detail, we bin the post-processed particle sub-sample as a function of density. Then, we iterate over all non-evolved particles, find the corresponding density bin, and identify the spatially closest evolved particle in the bin, finally associating its species abundances to the target particle.
    It is important to notice that, because of the intrinsic scatter in the particle density distribution resulting from the hydrodynamic evolution, the interpolation can be moderately affected by the particles chosen for the chemical evolution.
    In addition, the accuracy of the results will depend on the chosen interpolation scheme. For instance, a simple density-only interpolation (using the average abundances in each density bin) would be much cheaper, but at the expense of losing information about local effects, hence of worsening the accuracy of the interpolation scheme. Most importantly, both these approaches are valid as long as the gas is isothermal, whereas a temperature-dependent binning might be also necessary when the isothermality is not guaranteed (unless a 1:1 relation between density and temperature can be found).
    \item After each step, we create a new snapshot file of the simulation that also includes the species abundances, and we repeat the procedure from step (iv), using the resulting abundances as initial values for the new step. 
\end{enumerate} 
This procedure is applied up to the time when a sink particle is identified in the snapshot, in order to avoid any artificial effect that the removal of density can have on the chemistry calculations. We note, however, that the calculations could be performed even after sink particles formed, if the proper caveats are considered.
\label{sec:subset}
\begin{figure*}
    \centering
    \includegraphics[width=\textwidth]{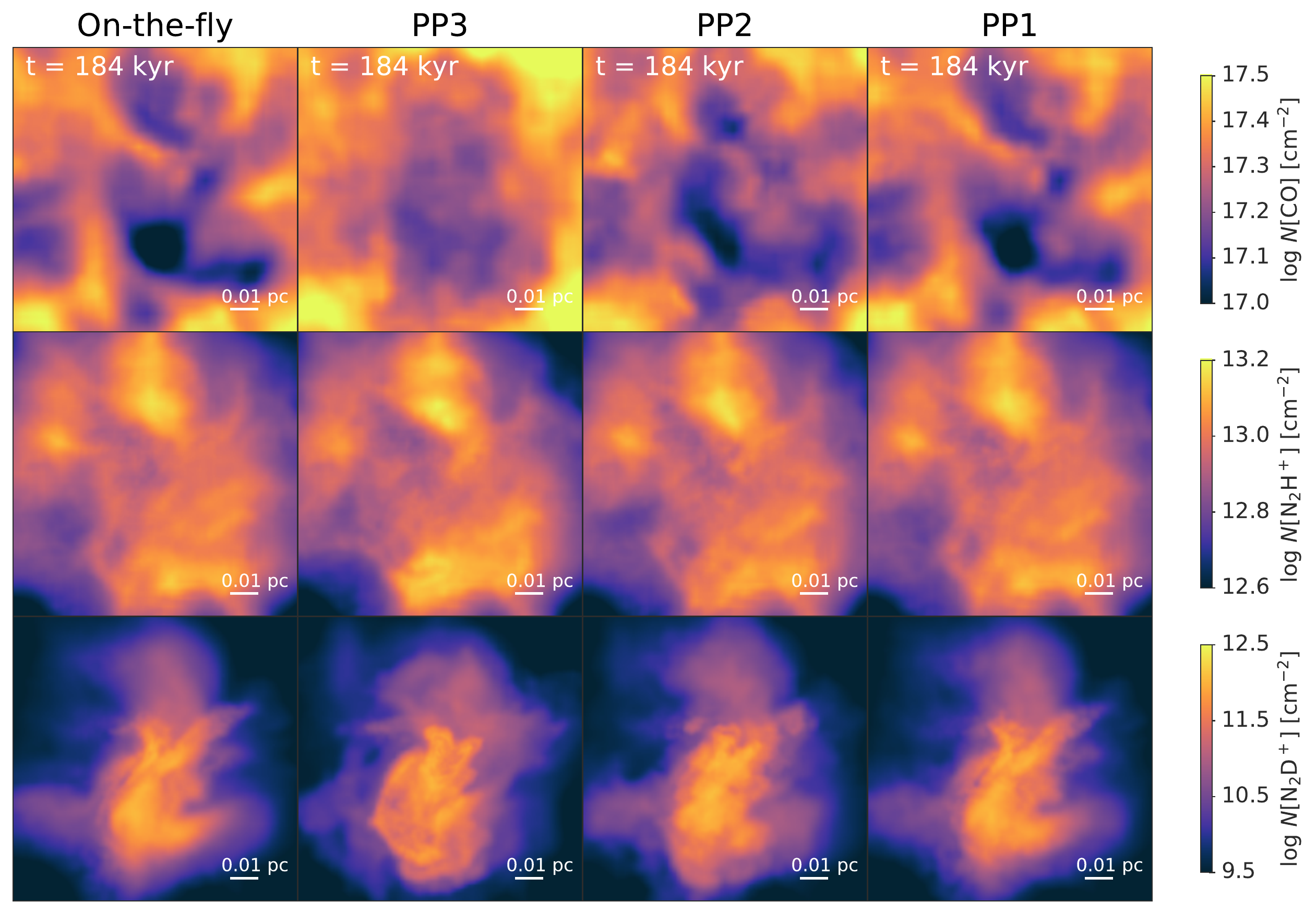}
    \caption{Column density maps of CO (first row), N$_2$H$^+$ (second row), and N$_2$D$^+$ (last row) at $t\sim184$ kyr for the `M1' core in \citet{Bovino2019}. The first column corresponds to the reference simulation (with OTF chemistry), and the second, third and fourth one to PP3, PP2, and PP1, respectively. While the agreement is reasonable in all cases, with the global distribution and density interval being always reproduced, the post-processing smears out some important features found in the full simulation (see, as an example, the ``filamentary'' structures in the top and bottom part of the N$_2$D$^+$ maps, or the location of the local minima in the CO maps), with the PP3 case showing the largest differences, as expected.}
     \label{fig:postproc_maps}
\end{figure*}

\section{Post-processing validation}
\label{sec:validation}
The approach considered in this work has two main free parameters: the fraction of particles used for the actual chemical evolution $f_{\rm part}$ and the time interval for the density update $\Delta t$. In this section, we validate the post-processing technique by applying it to the aforementioned M1 slow-collapse core run presented in \citet{Bovino2019}, employing the same network they used for consistency, and show how the free parameters affect our results.
\begin{figure*}
     \centering
     \includegraphics[width=.9\textwidth]{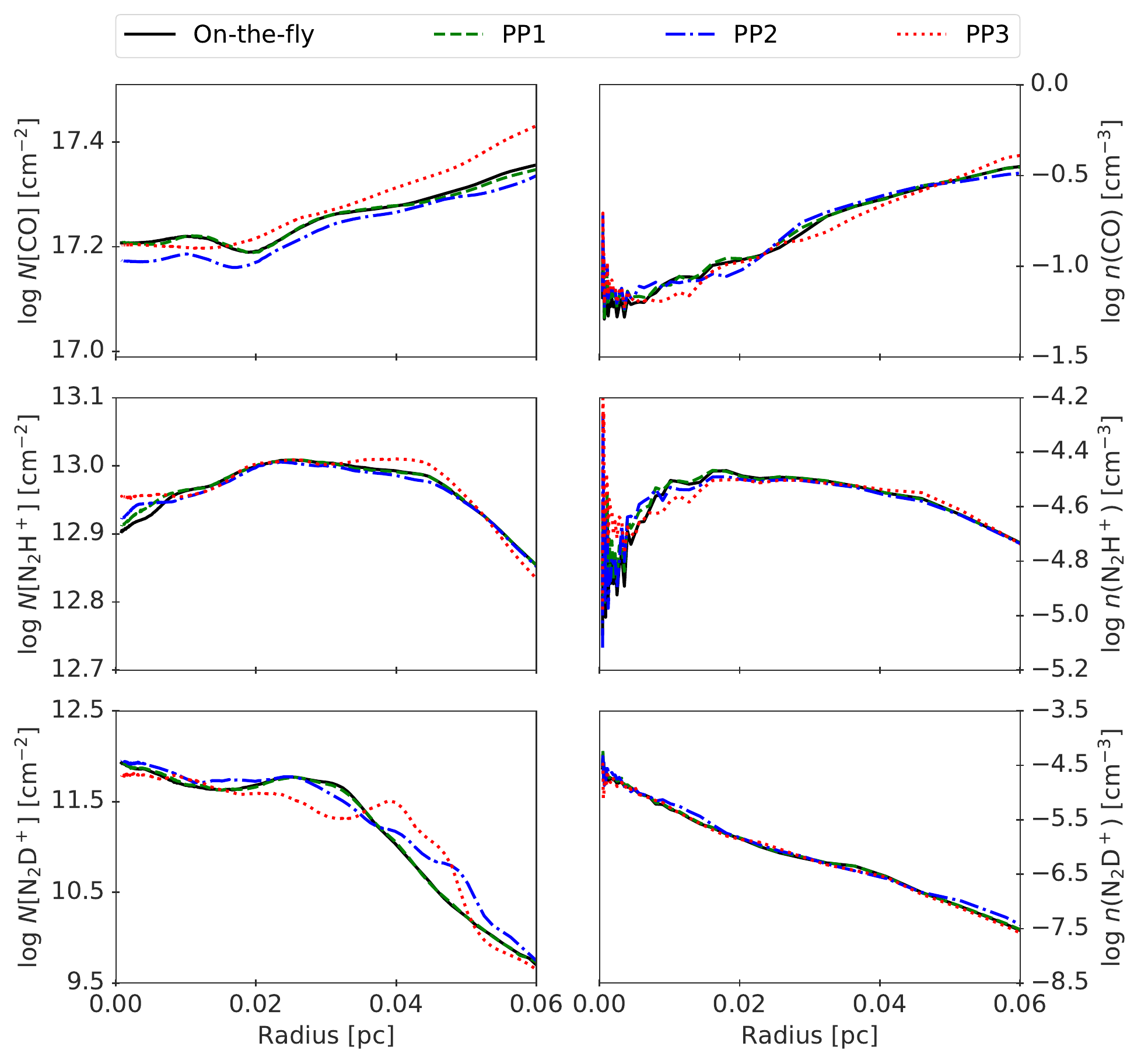}
     \caption{Radial profiles at $t\sim 180$ kyr of CO (top panels), N$_2$H$^+$ (middle panels), and N$_2$D$^+$ (bottom panels) column density (left column) and number density (right column), for the simulation with OTF chemistry (black solid lines), and post-processed chemistry, PP1 (using 10 percent of the particles;  green dashed lines), PP2 (using 1 percent; blue dot-dashed lines), and PP3 (using 0.1 percent; red dotted lines), respectively.}
     \label{fig:profiles_fpart}
\end{figure*}

\subsection{The sub-sample selection}
We start by considering how the fraction of particles taken from the simulation for the post-processing affects our results. We select three different post-processing (PP) cases, comparing them with our reference simulation: one taking 10 percent (PP1), another taking 1 percent (PP2), and a third taking 0.1 percent (PP3) of the simulation particles, respectively, and assume an integration time-step of 0.5~kyr, corresponding to the time interval between the snapshots in our hydrodynamic simulations, i.e. $\Delta t = \Delta t_0$. This means that, with the original snapshot having $6\times 10^5$ particles, the post-processing of these three validation cases is performed on 60000, 6000, and 600 particles, respectively. Furthermore, for all the analysis reported in the rest of this work, we assume $\xi_{\rm core}=0.5$.\footnote{We explored the effect of choosing different core-to-background particle ratios and, in this case, the differences were entirely negligible, both qualitatively and quantitatively.}

In Fig.~\ref{fig:postproc_maps}, we show a visual comparison of the CO, N$_2$H$^+$, and N$_2$D$^+$ column densities at $t\simeq180$~kyr. The MHD simulation with coupled on-the-fly (OTF) chemistry is shown in the first column, while the results of the post-processing are reported in the other columns: PP1 in the second one, PP2 in the third one, and PP3 one in the last one, respectively. Qualitatively, the global distribution is well recovered, but we notice that the post-processing is not always able to reproduce all the features in the simulation. 

For example, for N$_2$D$^+$ and N$_2$H$^+$ some artificial features in the maps tend to appear/disappear in the post-processing (e.g. filaments, overdensities, clumps etc.). On the contrary, for CO, which goes through the strongly time- and density-dependent freeze-out process, we observe an overabundance in the central region where we expect high depletion, which is progressively corrected as $f_{\rm part}$ increases.

To better quantify how important these differences are, we report in Fig.~\ref{fig:profiles_fpart} the column (left-hand panels) and number density (right-hand panels) radial profiles of the same three species shown in Fig.~\ref{fig:postproc_maps}. The column density profiles are obtained by radially averaging at set radii the correspondent column density map, while for the number density profiles we obtain the average over spherical shells of the actual 3D particle distribution. Evaluating the error on the number densities helps understanding how much uncertainty the integration along the line-of-sight is bringing, and to better assess the real error on local quantities, rather than averaged ones. Our reference run is shown as a black solid line, PP1 as a green dashed one, PP2 as a blue dot-dashed one, and PP3 as a red dotted one. The profiles clearly show that species depending mostly on gas-phase reactions (N$_2$H$^+$ in this case) are well reproduced, with low relative errors, even when only 1 percent of the particles is used. Species that heavily depend on gas-grain interactions, on the other hand, like N$_2$D$^+$ and CO, are more sensitive to the number of particles chosen for the post-processing, although the errors remain small overall.  
This might be especially relevant in regions where the density rapidly changes in time, on time-scales much shorter than $\Delta t_0$. Nevertheless, considering the huge dynamic range involved, we can conclude that the error is in general very small ($\lesssim 10\%$), hence making our post-processing method a viable and reasonably cheap alternative to on-the-fly calculations.

By looking more in detail at the differences between the number density and column density profiles, we can notice a similar behaviour in both number and column densities, that can be attributed to the incomplete tracing of the core density structure when using a low $f_{\rm part}$, that favours higher density regions, sampled with more particles. These errors are then consistently propagated to the column densities.

This, in general, means that the $f_{\rm part}$ choice does not directly affect the calculations of the chemical abundances per se, but using a low value will still amplify the errors because of the poor sampling of the structure under analysis. In addition, as already mentioned, the value of $f_{\rm part}$ does have a more important role on the task of reproducing ``observational'' features, with the species relying on gas-grain processes being slightly more affected in general that those controlled by mostly gas-phase reactions.

\subsection{Integration time effect}
\label{sec:timesteps}
\begin{figure*}
     \centering
     \includegraphics[width=.9\textwidth]{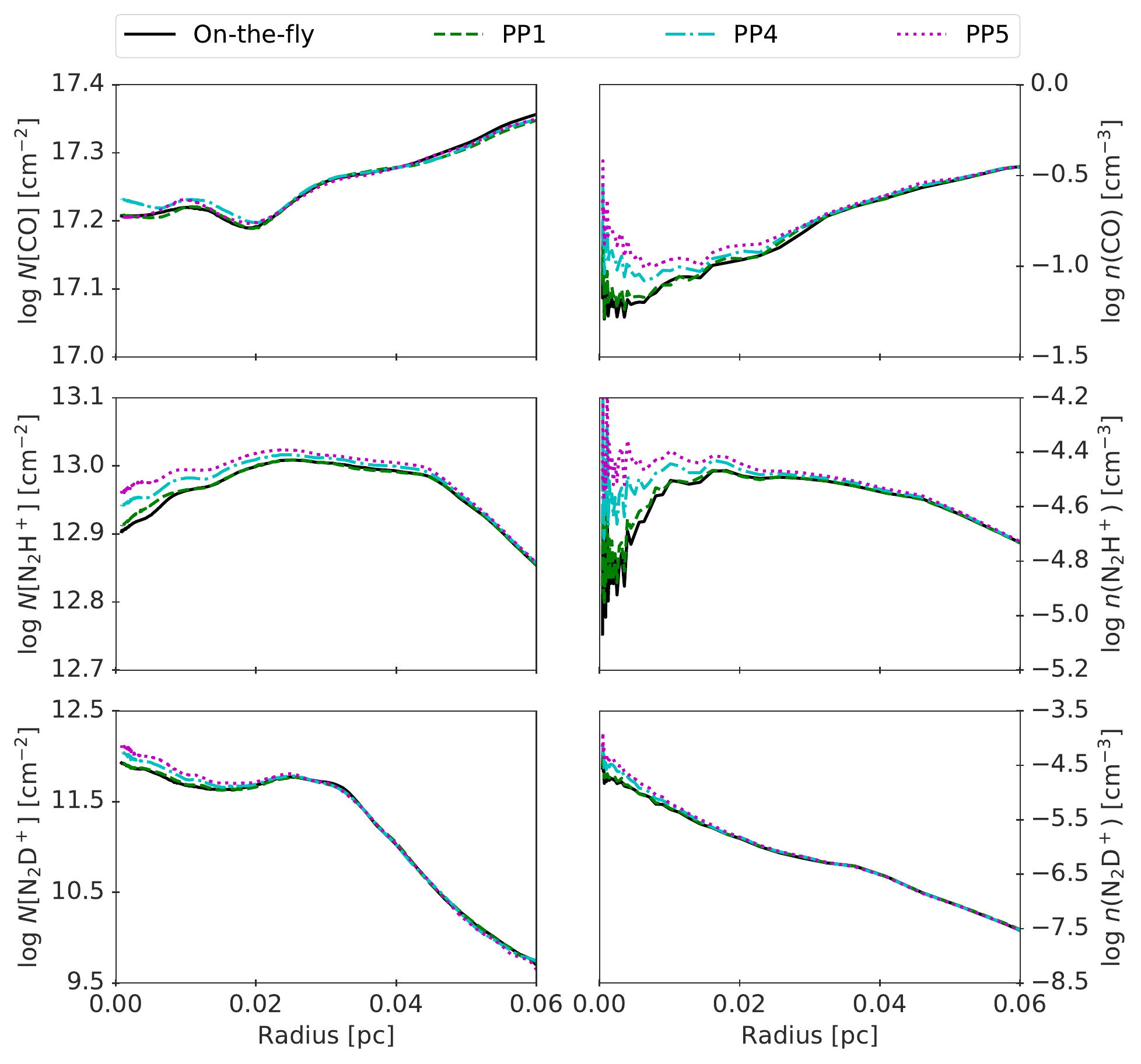}
     \caption{Same as Fig.~\ref{fig:profiles_fpart}, but now comparing the integration time-step choices. Here, the cyan dot-dashed line corresponds to the PP4 case, while the magenta dotted line corresponds to the PP5 case.}
     \label{fig:profiles_times}
\end{figure*}
The other important free parameter in our method is the time interval for the density update $\Delta t$. In the previous section, we assumed $\Delta t\equiv\Delta t_0 = 0.5$~kyr for simplicity, even though, in principle, $\Delta t$ might be arbitrarily set to larger values. However, we should keep in mind that, because of the dependence of chemical reactions on gas density, a different choice for the time-interval between density updates could have a significant impact on the evolution. In order to assess how sensitive the post-processing technique is to this choice, and thus its reliability, we analyse two additional cases, where we fix $f_{\rm part}=0.1$: a case with a five times longer integration step ($\Delta t=5\Delta t_0$; PP4), and a second one using an eight times longer one ($\Delta t=8\Delta t_0$; PP5). We stress that the density used at each time-step corresponds to the value obtained from the particle evolutionary history at the beginning of the time-step, in the aim at mimicking a coarser time sampling of the simulation outputs.

In Fig.~\ref{fig:profiles_times}, we show radial profiles of the number and column densities of the same species discussed before, comparing our reference run with OTF chemistry (black solid line) with PP1 (green dashed line), PP4 (cyan dot-dashed line), and PP5 (magenta dotted line). The right-hand panels, where we report the number density profiles, clearly show that the time-step choice heavily affects the high-density gas abundances of two of the three species shown (CO and N$_2$H$^+$), while at large radii (lower densities) the effect becomes negligible. In particular, when longer integration time-steps at constant density are used, the solution obtained for high-density gas, that is expected to evolve more rapidly, tends to approach the equilibrium solution, which does not necessarily match the solution in the reference simulation, leading to the observed overestimation. Interestingly, this effect is smeared out in the column densities  shown in the left-hand panels, because of the averaging effect resulting from the integration along the line-of-sight (LOS), and makes the time-step choice almost irrelevant. This result highlights how crucial a careful choice of the integration time-step is to improve the agreement with the reference simulation, and that it should be calibrated on the local quantities rather than on integrated ones.

\section{Error analysis and computational efficiency of the post-processing}
\label{sec:comparison}
\begin{figure*}
     \centering
     \includegraphics[width=\columnwidth]{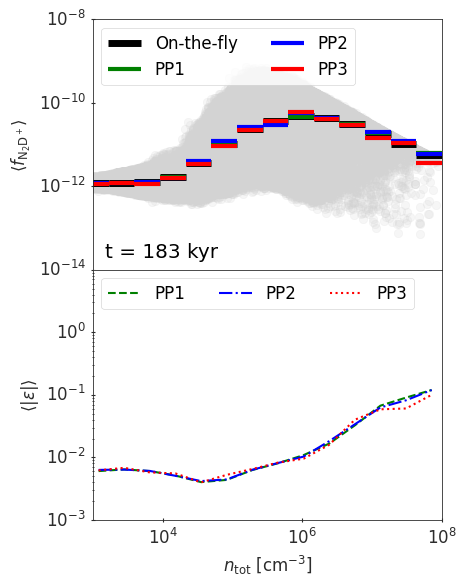}
     \includegraphics[width=\columnwidth]{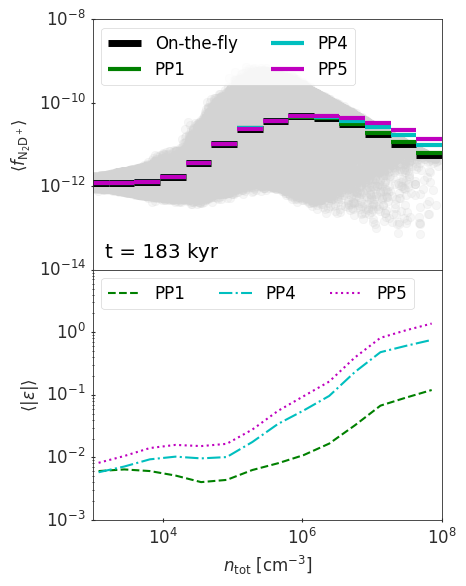}
     \caption{Fractional abundance of N$_2$D$^+$ as a function of density for our reference and PP runs at $t=183$~kyr. In the left panels, we show the impact of the $f_{\rm part}$ parameter, whereas in the right ones we show the effect of a longer integration time-step. The average abundance of N$_2$D$^+$ for the different runs is reported in the top panels, and the relative error between each PP run and our reference run in the bottom ones, following the same colour scheme of the previous figures. Finally, as a reference, in the top panels we also show the actual chemical abundance distribution from the reference run as a grey shaded area.}
     \label{fig:species}
 \end{figure*}
To better assess the observed differences in the validation processes, we now look more in detail at the error associated with the species abundance distribution. In Fig.~\ref{fig:species}, we focus in particular on the mass fraction of N$_2$D$^+$ ($f_{\rm N_2D^+}$) as a function of density, comparing the reference simulation with the post-processing at $t=183$~kyr.
It is important to mention that for all the subsequent analysis, we focus on  particles above a density threshold of $10^3$ cm$^{-3}$ only, in the aim at consistently tracing particles belonging to the core, removing any possible influence of the low density background particles on the results.
In the top panels, we report the average mass fractions directly obtained from the reference run (in black) and those obtained by the different post-processing cases, as described in Sections~\ref{sec:subset}~and~\ref{sec:timesteps}, in bins 0.4~dex wide. In the background, we also show the full particle distribution from the reference run as a grey shaded area. PP1 is shown in red, PP2 in blue, PP3 in green, PP4 in cyan, and PP5 in magenta. From the figure, it is clear that the agreement is extremely good in all cases, with deviations happening mostly in the density intervals where the abundance evolution is fast, hence the scatter large (i.e. $n\sim10^4-10^7\rm\, cm^{-3}$), where the post-processing is not always able to accurately follow the chemical evolution.
In the bottom panels, we show the average absolute value of the relative error, defined as
\begin{equation}
    \varepsilon\equiv \frac{f_{\rm N_2D^+,PP}-f_{\rm N_2D^+,OTF}}{f_{\rm N_2D^+,OTF}},
\end{equation}
where $f_{\rm N_2D^+,PP}$ is the mass fraction from the considered PP run and $f_{\rm N_2D^+,OTF}$ the one from the reference simulation with chemistry solved OTF. 
The line styles and colours are the same as in Figures~\ref{fig:profiles_fpart}~and~\ref{fig:profiles_times}, and correspond to results obtained with the post-processing for the evolved particles only (not the interpolated ones).
The impact of $f_{\rm part}$, reported in the left-hand panel, is almost negligible across the entire density range sampled, and the abundance errors never exceed the ten percent level. On the other hand, the integration time-interval has a stronger impact, as can be inferred from the larger variations at the highest densities, where the errors can reach higher values. In general, the discrepancies between the post-processing and the OTF calculations can be easily explained with the density evolution time-scale. In fact, when the density changes slowly, the constant density assumption over the time-step between simulation outputs is reasonable, with the abundances approaching the chemical equilibrium solution. When the density changes quickly, instead, the gas is not able to approach the equilibrium conditions, and the post-processing, that does not keep up with this evolution, gives inaccurate results.

To summarise, we report in Table~\ref{tab:errors} an estimate of the typical error of the post-processing technique, defined by the 10-th ($\varepsilon_{10}$) and 90-th ($\varepsilon_{90}$) percentiles of the relative error distribution for each post-processing case, resulting either from the post-processed particles only (fourth column), or from the whole particle set (last column). Also in this case, we only consider N$_2$D$^+$, for consistency with Fig.~\ref{fig:species}. The extremely small variation among PP1, PP2, and PP3 confirms that the post-processing is only moderately affected by the sampled fraction of particles chosen, with the subsequent interpolation scheme to recover the whole particle distribution being the main source of error, especially when a small fraction of particles is used, raising it to about a factor of $\sim2$ at most. On the other hand, for the runs with longer time-steps between chemistry calculations (PP4 and PP5), the typical error gets progressively worse both on the evolved particle subset and the whole distribution, as expected from the key dependence of the calculation of chemical abundances on the time evolution of the system.
\begin{table}
	\centering
	\caption{Typical error of the PP runs for N$_2$D$^+$ for the directly post-processed particles only ($\varepsilon_x^{\rm e}$) and the entire particle distribution (evolved and interpolated; $\varepsilon_x^{\rm t}$), with $x=10$ or 90 the error distribution percentiles (see text). Consistently with the previous analysis, the impact of $f_{\rm part}$ is very moderate, whereas the time interval choice has a much stronger impact.}
	\label{tab:errors}
	\begin{tabular}{lccccc}
		\hline
		Model & $f_{\rm part}$ & $\Delta t$ & $[-\varepsilon_{10}^{\rm e};\varepsilon_{90}^{\rm e}]$ & $[-\varepsilon_{10}^{\rm t};\varepsilon_{90}^{\rm t}]$\\
		\hline
		PP1 & 0.1 & $\Delta t_0$ & $[<0.001;0.013]$ & [0.259;0.384] \\
		\\
		PP2 & 0.01 & $\Delta t_0$ & $[<0.001;0.014]$ & [0.476;1.139] \\
		PP3 & 0.001 & $\Delta t_0$ & $[<0.001;0.013]$ & [0.667;1.888] \\
        \\
        PP4 & 0.1 & $5\Delta t_0$ & [0.012;0.059] & [0.254;0.489] \\
		PP5 & 0.1 & $8\Delta t_0$ & [0.023;0.100] & [0.248;0.569] \\
		\hline
	\end{tabular}
\end{table}

In conclusion, our analysis suggests that, in order to get a good accuracy in the chemical evolution, a high output frequency of the underlying hydrodynamic simulation (without chemistry) is needed to avoid large errors in the recovered abundances. In particular, our choice for $\Delta t_0$ was dictated by the simultaneous need of a high output frequency (able to accurately track the typical dynamical evolution time-scales in the simulation) and a low enough number of snapshots (given the large storage overhead due to the chemical species abundances and the long timescales we wanted the evolution to last). Although our choice has been shown to be adequate, an even higher output rate should be preferred when chemistry is not solved OTF, as is the case for the runs aimed at be post-processed. Note that, however,  this would make the post-processing approach more computationally expensive, hence less convenient against a fully coupled simulation. When chemistry is solved on-the-fly, instead, such a high output frequency is not needed, since chemistry is consistently evolved on time-scales that are typically much shorter than those between outputs.

\begin{figure}
    \centering
    \includegraphics[width=\columnwidth]{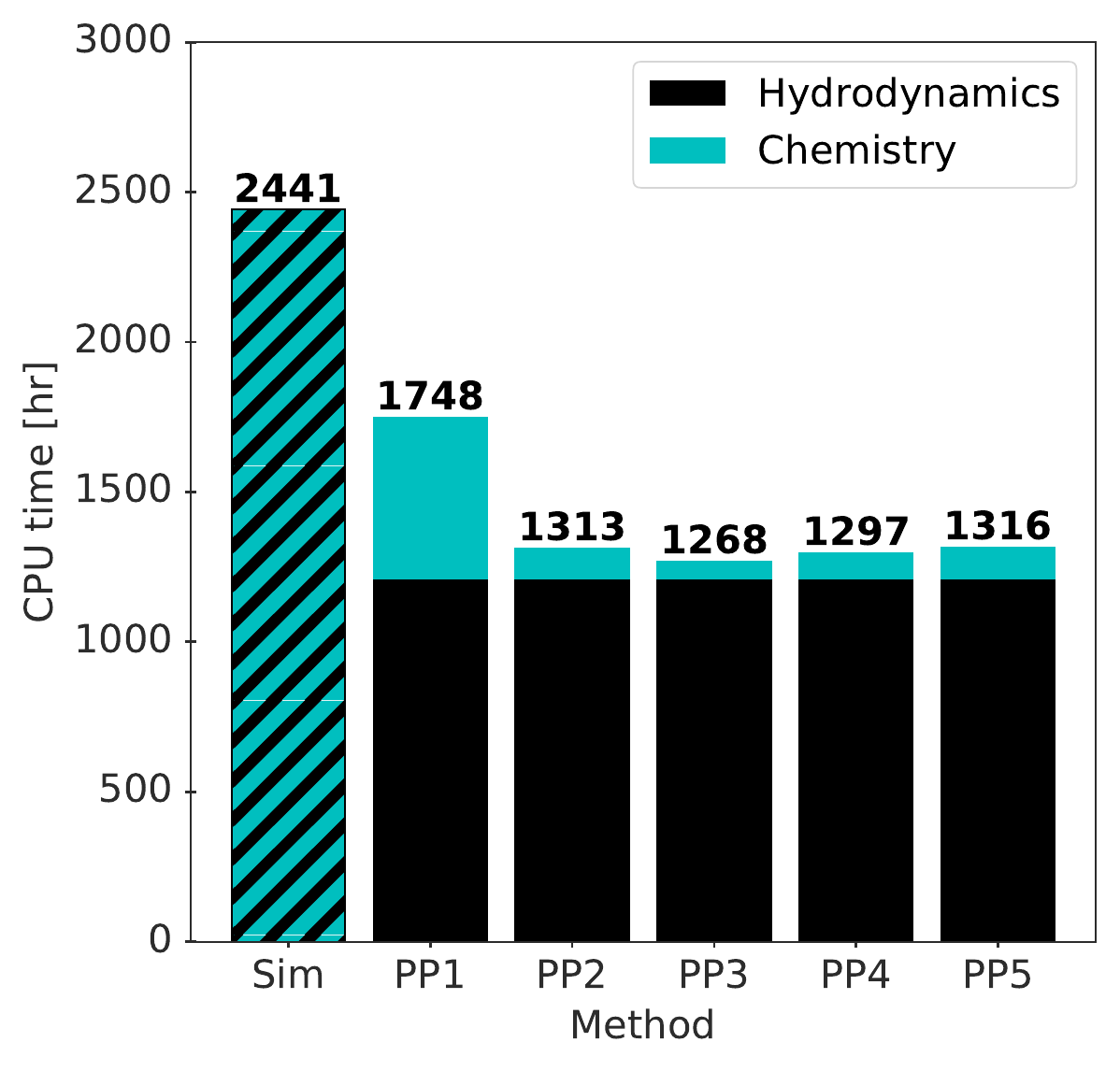}
    \caption{CPU time in hours used for the reference simulation with OTF chemistry (Sim), and the different post-processing cases, PP1 ($f_{\rm part}=0.1$), PP2 ($f_{\rm part}=0.01$), PP3 ($f_{\rm part}=0.001$), PP4 ($\Delta t = 5 \Delta t_0$), and PP5 ($\Delta t = 8 \Delta t_0$).}
    \label{fig:postproc_cputime}
\end{figure}

Finally, a last aspect of our post-processing method needs to be addressed, i.e. its computational efficiency, that has been improved in our analysis via OpenMP parallelisation. In Fig.~\ref{fig:postproc_cputime}, we report the computational cost of the different cases as the total time of the run in CPU hours, to highlight how much time is gained via the post-processing technique relative to OTF calculations. Here, our reference MHD simulation with OTF chemistry is shown as a black/cyan bar, whereas the post-processing models are split in a black part, the same for all models, corresponding to the cost of a MHD run without OTF chemistry, and an cyan part resulting from the individual cost of the post-processing step, i.e. evolution+interpolation. As expected, the cost increases in a sub-linear fashion with the number of particles in the post-processed sub-sample, because of the competing effect of the actual post-processing and the interpolation step.
Obviously, if we extrapolate the CPU time to all the particles of the simulation, we expect to obtain a cost comparable to that of the fully coupled simulation, hence the post-processing stops being convenient. If we instead increase the time interval between density updates, the cost mildly decreases, resulting in no significant gain. This is due to the fact that, although the number of density updates is reduced, the chemical calculations must be performed over a longer time-scale per step, resulting in only moderately different computational times. Moreover, as discussed above, longer time-steps can significantly overestimate the chemical abundances, especially at the highest densities, making this choice generally less convenient.

Concluding, our analysis suggests that the post-processing technique applied to small-scale simulations yields good results, with relative errors on average at the percent level being achieved by using a subset of just 10 percent of the total particles, and assuming short enough, but still reasonable, time-steps between density updates. The method appears reliable also when the number of particles used is significantly smaller than that of the simulation ($f_{\rm part}\lesssim 0.1$), and its accuracy is not affected by this choice, as long as the the time interval between density updates is not increased too much. Nevertheless, in our experiments, the actual gain of the post-processing compared to coupled OTF chemistry calculations is not more than a factor of a few, hence it appears convenient only when the same simulation is expected to be used multiple times to explore the relative importance and the effect of different chemical processes, rates, or when including huge networks, which are challenging to solve on the fly. As a final note, based on our findings, and on the average free-fall time $t_{\rm ff}$ of the considered core, we can speculate that an output frequency of $\Delta t\ll 10^{-2}t_{\rm ff}$ is sufficient to maintain the chemical abundances of the majority (80\%) of the post-processed particles below an average error of 1\%.

\section{Conclusions}
\label{sec:conclusions}
In this study, we assess the reliability and the uncertainties of the chemical post-processing technique applied on 3D MHD simulations, with the aim at facilitating the implementation of complex chemical networks in the context of star and planet formation processes. For our analysis, we employed one of the simulations of collapsing massive cloud fragments introduced by \citet{Bovino2019}. After having extracted the dynamical evolution of a subset of particles, we evolved the chemistry via the non-equilibrium chemical code \textsc{krome}, and interpolated the abundance for the remaining particles in each simulation output to recover an approximate version of the chemical abundances for the full particle distribution. To validate such an approach, we compared the results obtained with different combinations of key input parameters, namely, the fraction of particles extracted for the post-processing and the coarse time-step used for the chemical calculations, against the simulations performed by \citet{Bovino2019}, in which the chemistry calculations were performed on-the-fly fully coupled within the MHD code \textsc{gizmo}. This is the first time that the effective errors related to post-processing procedures is comprehensively explored, made it possible from our powerful 3D MHD simulations with on-the-fly non-equilibrium chemistry.  

Overall, we found that the post-processing technique is quantitatively reliable, provided the particle sub-sample to be directly post-processed is, at least, above 1\% of the total simulation ensemble, in order to limit the impact of the interpolation scheme on the abundance error, and significantly cheaper than simulations with fully coupled OTF chemistry. However, if the aim is to recover specific features as those obtained in the MHD simulations with OTF chemistry, a larger amount of post-processed particles ($f_{\rm part} \geq 0.1$) must be used, resulting in a less significant gain of computational time. Although our results suggest that the exact choice of the parameter $f_{\rm part}$ only moderately affects the results, we stress that, when the particle sampling is not good enough (below 1\%), the chemical evolution will more coarsely sample the fiducial solution, independent of the chosen time-step, and the results will not always trace the chemical structure of the system under study.

A few caveats in the post-processing approach must be mentioned: 
\begin{itemize}
\item First, a crucial issue is the interpolation of the non-post-processed particles, which is based on the density-binned abundances obtained from the post-processed particles. While the coupling of the density-binned interpolation with a distance-based neighbour selection to assign the species abundances allows to take into account the evolutionary history of the gas around the target particle, it is still not sufficient to perfectly recover the abundances, thus introducing some artefacts in the species abundances. However, we have to keep in mind that the intrinsic scatter can only be recovered with OTF chemistry, because of its strong dependence on particle-by-particle variations in the dynamic properties, typically occurring on timescales that cannot be easily tracked with the post-processing unless an extremely short output time-step is assumed (removing any gain of the post-processing). In addition, compared to a simpler density-based-only interpolation, in which the spatial localisation of the gas particles is completely neglected, and the scatter in the chemical abundances completely washed out, our density+proximity approach is able to improve the accuracy of the post-processing by a factor of 2--3.
\item An additional issue, which is connected to the one just discussed, is the poor sampling of some density intervals, that would produce large errors also in the average chemical properties. This limitation can be overcome by appropriately tuning the $\xi_{\rm core}$ parameter, and increasing $f_{\rm part}$, but at the cost of a more demanding analysis. 
\item It is also important to notice that a low number of particles does not allow to recover possibly important spatial features in the system, and should be avoided, in particular considering the moderate increase in computational cost when a larger fraction of particles is employed.
\item Finally, a fundamental limitation of the post-processing approach is in general the time-step over which the density is kept fixed for the chemistry calculations. While a higher output frequency can ideally improve the accuracy of the results, it also makes the post-processing more expensive, hence closer to a proper OTF chemical evolution. 
\end{itemize}

Apart from these caveats, the post-processing of simulations with the method described in the present work can be extremely useful when exploring the relative effect of key input parameters, saving a considerable amount of time and (super-)computing resources, or when it is necessary to employ large networks that would be otherwise impossible to include on-the-fly. However, when a very accurate model as close to reality as possible is required (within the limits of the underlying chemical model), on-the-fly chemical calculations within (magneto-)hydrodynamic simulations should always be preferred, especially when there is a direct feedback between the evolution of chemistry and dynamics, e.g. when the dynamical evolution of the gas is also determined by the temperature variations, that in turn are properly traced by the balance between chemical reactions.

\section*{Acknowledgements}
We thank the anonymous referee for the useful suggestions that improved the quality of the manuscript.
SFC thanks funding from CONICYT Programa de Astronom\'ia Fondo ALMA-CONICYT 2017 Project \#31170002.
AL acknowledges support from MIUR under the grant PRIN 2017-MB8AEZ.
SB acknowledges financial support from Millenium Nucleus NCN19\_058 (TITANs), and BASAL Centro de Astrofisica y Tecnologias Afines (CATA) AFB-17002. The simulations were performed with resources provided by the KULTRUN Astronomy Hybrid Cluster at Universidad de Concepción.

\section*{Data Availability Statement}
The data underlying this article will be shared on reasonable request to the corresponding author.


\bibliographystyle{mnras}
\bibliography{postproc} 




\appendix


\bsp	
\label{lastpage}
\end{document}